\documentclass{elsart}
\usepackage{natbib}
\usepackage{psfig}
\newcommand{\asca}{{\sl ASCA} }
\newcommand{\sax}{{\sl SAX}}

\newcommand{\euve}{{\sl EUVE} }

\begin{document}
\runauthor{Takahashi, Madejski, and Kubo}
\begin{frontmatter}

\title{X--ray Observations of TeV Blazars and Multi-Frequency Analysis}

\author[ISAS]{Tadayuki Takahashi,}
\author[GSFC]{Greg Madejski,}
\author[TIT]{and Hidetoshi Kubo}

\address[ISAS]{Institute of Space and Astronautical Science, Sagamihara, Kanagawa 229-8510, Japan}
\address[GSFC]{Lab for High Energy Astrophysics, Code 662,
NASA/GSFC, MD 20771, USA}
\address[TIT]{Department of Physics, Tokyo Institute of Technology, Meguro-ku, Tokyo 152-8551, Japan}

\begin{abstract}

The non-thermal spectra of blazars, observed from radio to GeV/TeV
$\gamma$--rays, reveal two pronounced components, both produced by 
radiation by energetic particles.  One peaks in the IR - to soft X--ray band, 
radiating via the synchrotron process;  the other, peaking in the 
high-energy $\gamma$--rays, is produced by the Compton process. 
These spectra -- and, in particular, the \asca data -- suggest that the 
origin of the seed photons for Comptonization is diverse.  In the 
High-energy peaked BL Lac objects (HBLs), the dominant seed photons 
for Comptonization appear to be the synchrotron photons internal to 
the jet (SSC process).  In the quasar-hosted blazars (QHBs), on the other 
hand, the X--ray band emission is still dominated by the SSC process, while 
the MeV to GeV range is produced by Comptonization of external photons 
such as the emission line light.  In the context of this three-component 
model, we derive the magnetic field of 0.1 - 1 Gauss for all classes of
blazars.  Lorentz factors $\gamma_{peak}$ of electrons radiating at each 
peak of the $\nu F(\nu)$ spectra are estimated to be $\sim
10^{5}$ for HBLs; this is much higher than $\sim 10^{3}$ for QHBs.  This
difference is consistent with the fact that the four sources that are
known to emit TeV $\gamma$--rays (TeV blazars) are all classified as
HBLs.  Among the TeV blazars, Mkn 421 is one of the brightest and most
variable emitters from ultraviolet (eV) to hard $\gamma$--ray (TeV)
energies, and its correlated inter-band variability suggests that both keV
and TeV spectral regimes are produced by the same, most energetic end
of the electron population radiating via the synchrotron process in
the keV, and the SSC process in TeV band.  The
multi-frequency observations including TeV energy band provide the 
best opportunity to understand high-energy emission from blazar jets.
In this paper, we discuss results of multi-frequency analysis and 
review the results of intensive campaigns for Mkn 421 from 1994 to 1998.

\end{abstract}
\end{frontmatter}

\section{Introduction}

The launch of the Compton Gamma-Ray Observatory and the major
improvements to ground-based Cherenkov telescopes marked the 1990s as
the pivotal years in the field of astrophysics at the very highest
energies.  In particular, the detection of GeV $\gamma$--rays from
over 50 blazars by the EGRET telescope on CGRO (cf. Mukherjee et al. 
1997) established them as a class of extreme ``particle accelerators''
-- on two accounts: they produce the most energetic photons of any known 
extragalactic cosmic sources, but, assuming isotropic emission, they are 
also the most luminous.  It is generally thought that the non-thermal emission
from blazars, observed in virtually every band from radio to GeV/TeV
$\gamma$--rays, is produced by very energetic particles via both
synchrotron and Compton processes.  Most of these objects are variable, 
showing flares with the greatest amplitude at the highest energies of 
each component, on time scales as short as a day.  This underlines the 
importance of the high-energy emission for any modeling.  Via the 
$\gamma$--ray opacity arguments (see, e.g., Mattox et al. 1993), 
this suggests that the $\gamma$--ray emission is produced in a very 
compact region, moving at relativistic speeds at a small angle towards 
the observer, most likely in a form of a jet. 

Multi-frequency studies of blazars reveal that the overall spectra have at
least two pronounced components: the low energy peak (LE) and the high
energy peak (HE) (see, e.g., von Montigny et al. 1995). For the blazars
showing quasar-like emission lines (QHBs), and for BL Lac objects
discovered via radio-selection techniques (the Low-Energy Peaked BL Lac
objects, or LBLs), the LE component peaks in the infrared, while in the BL
Lacs found using X--rays (the High-Energy Peaked BL Lac objects, or HBLs),
it peaks in the ultraviolet or even in soft X--rays (see, e.g., Sambruna,
Maraschi, \& Urry 1996). The HE component, on the other hand, peaks in the
$\gamma$--ray band, in the MeV - to - GeV range, and in the case of a few
HBLs, it extends to the TeV range. 

The local power-law shape, the smooth connection of the entire radio -
to - UV (and, for the HBLs, soft X--ray) continuum, as well as the
relatively high level of polarization observed from radio to the UV,
implies that the emission from the LE component is most likely
produced via the synchrotron process by relativistic particles
radiating in magnetic field.  The HE component is then believed to be
produced via Comptonization by the same particles that radiate the LE
component.  The source of the ``seed'' photons can be the
synchrotron radiation internal to the jet -- as in the
Synchrotron-Self-Compton (SSC) models (Rees 1967; Blandford \& Konigl
1979; Konigl 1981; Ghisellini \& Maraschi 1989, Inoue \& Takahara 1996).  Alternatively,
these photons can be external to the jet, as in the External Radiation Compton
(ERC) models: either the UV accretion disk photons (Dermer,
Schlickeiser, \& Mastichiadis 1992), or these UV photons reprocessed
by the emission line clouds and/or inter-cloud medium (Sikora, Begelman, 
\& Rees 1994; Blandford \& Levinson 1995), or else, IR radiation ambient 
to the host galaxy (Sikora et al. 1994).  The X--ray regime is important, 
as it is where the emission due to both processes overlap: in the context 
of the above scenario, in HBLs, X--rays form the high energy tail of the
synchrotron emission, while for QHBs, they form the lowest observable
energy end of the Comptonized spectrum.

The discovery of the TeV emission from four nearby BL Lac objects (Mkn
421, Mkn 501, 1ES2344+514, and PKS 2155-304) by ground based Cherenkov
telescopes (Punch et al. 1992;  Quinn et al. 1996;  Catanese et al. 1998;  
Chadwick et al. 1998) has given us an opportunity to study the radiation 
processes responsible for production of photons at such extreme energies.  
The first multi-frequency observation from Mkn 421 from radio to TeV
$\gamma$--rays, conducted in 1994, revealed that while the keV X--ray
and TeV $\gamma$--ray fluxes varied nearly simultaneously, the flux in
other bands remained relatively steady (Macomb et al. 1995;  Takahashi
et al. 1996a).  Subsequent multi-frequency campaigns for this object also 
revealed correlated flare activity in X--rays and TeV $\gamma$--rays, 
suggesting that both keV and TeV spectral regimes are produced by the
same, most energetic end of an electron population, radiating via the 
synchrotron process in the keV band, and Compton process in the TeV bands.  
Given the relative absence of emission lines in the TeV - emitting HBLs, the 
``seed'' photons for Comptonization are most likely internal to the jet, as 
in the SSC models.  

The multi-frequency spectra of HBLs Mkn 501 and PKS 2155-304 are similar
to that of Mkn 421:  the LE components of these objects peak in the X--ray
band.  Mkn 501 showed a remarkable flare activity in April 1997. During the
multi-frequency campaign, both X--ray and TeV $\gamma$--rays increased
by more than one order of magnitude from quiescent level (Catanese et
al. 1997; Pian et al. 1998; Kataoka et al. 1999), suggesting 
that the same mechanism takes
place as in the case of Mkn 421. However, in Mkn 501, the synchrotron
emission in quiescence peaked below 0.1 keV, while during the 1997 flare, it
peaked at $\sim$ 100 keV.  This is in contrast to Mkn 421, 
where the position of the LE peak shows relatively little change.  
For PKS 2155-304, one of the best studied BL Lac objects in other bands, 
the TeV emission was detected only recently (Chadwick et al. 1998), 
most likely due to its Southern location, and a relative lack of TeV 
telescopes in the Southern hemisphere.  PKS 2155-304 had shown very 
different behavior in two subsequent campaigns (Brinkmann et al. 1994; 
Urry et al. 1997;  Ghisellini et al. 1998).

Here, we report the results of \asca observation of blazars in the context 
of their multi-frequency emission. Based on the analysis of multi-frequency 
spectra, we discuss characteristics of TeV blazars.  We then review results 
obtained from our extensive campaigns of one of the brightest TeV blazars, 
Mkn 421.

\section{ASCA Observations of Blazars}

{\sl ASCA} observed 18 blazars, of which 10 were also observed
contemporaneously with $EGRET$ as parts of multi-wavelength campaigns, 
and these show a clear difference in the spectra and variability
between HBLs, LBLs, and QHBs.  We constructed the multi-frequency
spectral energy distributions for these 18 sources and performed
multi-frequency analysis to study fundamental parameters in the
emission from blazars (Kubo 1997; Takahashi et al. 1997; Kubo et al. 
1998, hereafter KTM98).

Two examples of blazar spectra plotted in $\nu F(\nu)$ representation
(giving the emitted power per decade of energy under an assumption of 
isotropic emission) are shown in Figure 1.  One is of Mkn 421 (Macomb 
et al. 1995;  Takahashi et al. 1996a), a fairly typical HBL; its X--ray 
spectrum is soft (steep) and on the extrapolation of the ultraviolet, 
suggesting that the synchrotron emission extends into the X--ray range.  
The other is PKS 0528+134 (KTM98), a strong lined QHB; in contrast to 
Mkn 421, the X--ray spectrum of PKS 0528+134 is harder than the UV spectrum,
implying that the X--rays form the onset of the second, HE component. 

The \asca X--ray spectra of HBLs are the softest, with the power law
energy index $\alpha \sim 1 - 2$.  The X--ray spectra of the QHBs are
the hardest ($\alpha \sim 0.6$).  For LBLs, the spectra are
intermediate; in one case, 0716+714, the spectrum shows hardening with
an increasing energy. The differences in the spectra between
subclasses were also observed with $ROSAT$ in the soft X--ray band
(Sambruna et al. 1996; Urry et al. 1997; Sambruna 1997; Padovani,
Giommi, \& Fiore 1997; Comastri et al. 1997). While for HBLs, the 
spectral index often varies with intensity on a time scale shorter than 
a week, for QHBs, it remains almost constant when the flux changes.

\begin{figure}[bt!]
\centerline{\psfig{figure=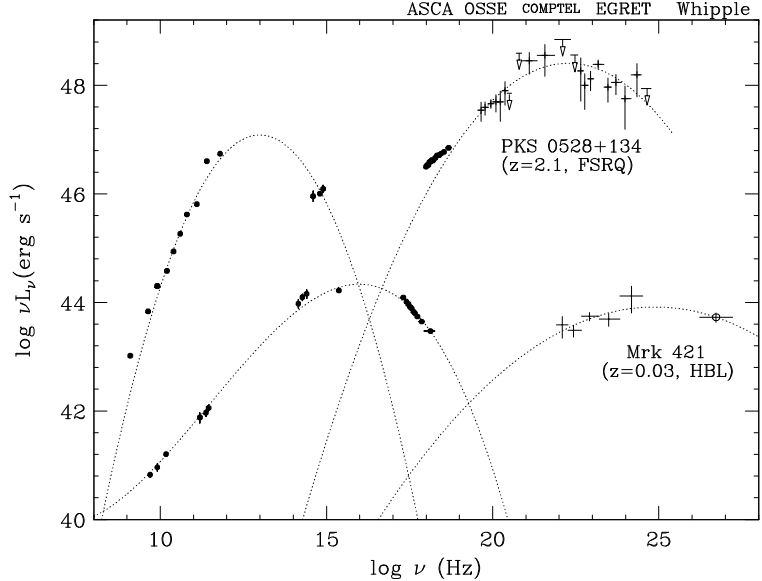,width=10cm}}
\caption{The multi-frequency spectrum obtained from the simultaneous 
observations of Mkn 421 and PKS 0528+134. The dotted
line shows the results of fit to a third-order polynomial function for 
low and high energy components. }
\end{figure}

\section{Multi-frequency Analysis - Magnetic Field and Electron Lorentz 
Factors} 

In the very broadest sense, the overall spectra of blazars can be 
characterized by four parameters:  the peak frequency of the synchrotron 
and Compton components ($\nu_{sync}$ and $\nu_{Comp}$), and luminosity of each 
component ($L_{sync}$ and $L_{Comp}$, calculated in the observer's frame, 
and assuming isotropy of emission).  Fig.~2 shows the distribution of the 
ratio $L_{Comp}$/$L_{sync}$ as a function of the peak of the LE component 
of the blazars considered by KTM98.  In QHBs, the ratio is much higher 
than that obtained from HBLs.  In the context of the synchrotron 
model, $\nu_{sync}$ and $L_{sync}$ are determined by the intensity of 
the magnetic field and the distribution function of electron energies.  
Similarly, in the context of the Compton model, $\nu_{Comp}$ and $L_{Comp}$ 
are related to the distribution functions of electron and target photon 
energies.  Below, we summarize a formalism useful for the determination of 
the magnetic field $B$ and electron Lorentz factors $\gamma_{el}$, 
following that presented in Takahashi et al. (1996a) and in KTM98.  

When the radiation is due to a single population of relativistic electrons 
with a broken power law distribution of $\gamma_{el}$ and a break at 
$\gamma_{peak}$, $\nu_{sync}$ (as measured in the observer's frame) 
is given as:
\begin{equation}
\nu_{sync} = 1.2 \times 10^{6} \gamma_{peak}^{2} B \frac{\delta}{(1+z)} \quad {\rm (Hz).} 
\label{eqn1}
\end{equation}
Here $B$ is magnetic field in Gauss, measured in the comoving frame, and 
$\delta $ is the ``beaming'' (Doppler) factor defined as 
$\delta  = \Gamma_{j}^{-1} (1 - \beta \cos \theta)^{-1}$.  

In the case when the photons internal to the jet dominate the radiative 
energy density (such as in HBLs), $L_{Comp} = L_{SSC}$ and 
$\nu_{Comp} = \nu_{SSC}$.  
If the electron energy is still in the Thomson regime, 
($\gamma_{el} \times h\nu_{sync} << m_ec^2$),
the expected peak of the SSC component is 
$\nu_{SSC} = 4 \gamma_{peak}^{2} \nu_{sync}/3$.
The ratio of $L_{SSC}$ to $L_{sync}$ is then determined by the intensity 
of magnetic field, and the distribution function of electron energies;  it 
can be expressed as: 
$\frac{L_{SSC}}{L_{sync}}=\frac{u_{sync}}{u_{B}}$,
where the $u_{sync}=L_{sync}/(4\pi R^{2}c\delta^{4})$ is the 
energy density of the synchrotron photons, and 
$u_{B}=B^{2}/(8\pi)$ is magnetic field energy density.  
The beaming factor ($\delta$) is then given from above equations: 
\begin{equation}
\delta^2 = 1.6\times 10^{12} \frac{L_{sync}}{c R^2} 
\left(\frac{L_{sync}}{L_{SSC}}\right) \frac{\nu_{SSC}^2}{\nu_{sync}^4} \frac{1}{(1+z)^2}
\label{eqn3}
\end{equation} 

An application of the synchrotron self-Compton (SSC) model to the overall
spectral distribution and variability data of 18 blazars observed by \asca
implies that for the HBLs, the SSC model can indeed explain all available
data quite well.  Importantly, at least for Mkn 421, the parameters derived
from it are in close agreement with those independently inferred from the
spectral variability observed in the X--ray band by \asca (see Sec. 4.1 
below).  The model implies relativistic Doppler factors $\delta$ in the 
range of 5 -- 20, consistent with those derived from the VLBI data and from 
the limits inferred from $\gamma$--ray opacity to pair production,
$\gamma\gamma\rightarrow e^{+}e^{-}$. 

\begin{figure}
\centerline{\psfig{file=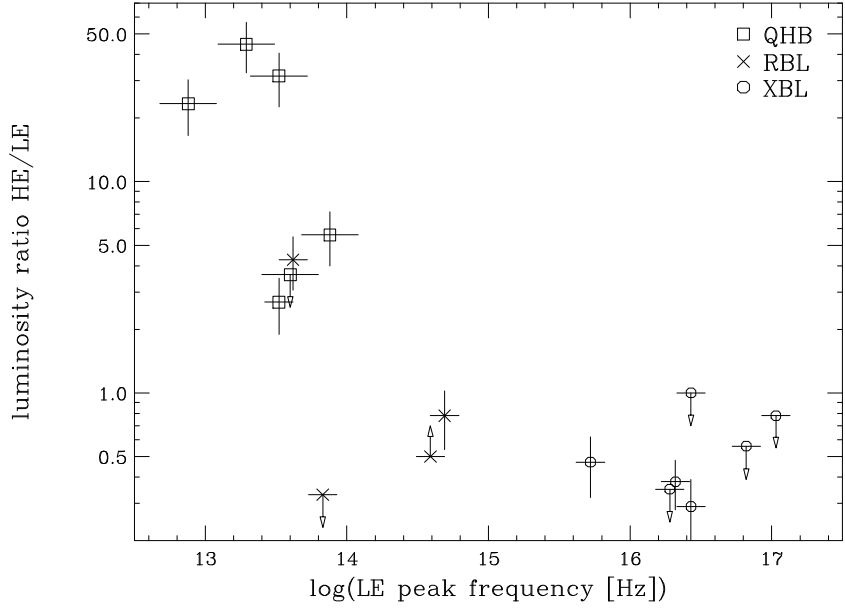,height=8cm}}
\caption{Distribution of the ratio $L_{Comp}$/$L_{synch}$ as a function of
the peak frequency of the LE component. Arrows indicate the upper or lower 
limits measured with $\gamma$--ray observations.}
\end{figure}

The situation in QHBs (and in some LBLs) is quite different.
In the X--ray band observed by \asca 
(0.7 -- 10 keV), the QHBs have spectra that are hard, with $\alpha
\sim 0.6$, and which are {\sl not} located on the extrapolation of the
synchrotron optical / UV spectra, but also apparently disjoint from the GeV HE 
component (cf. PKS 0528+134 in Fig.~1), suggesting that X--rays are likely 
to be due to a separate spectral component from GeV $\gamma$--rays. 
 The
application of the SSC model (thus assuming $L_{HE}$ is solely due to 
the SSC component) implies that the values of $\delta$ derived from 
Eq. 2 are much in excess of values inferred from the $\gamma$--ray opacity 
arguments or the VLBI data (for details, see KTM98).  This discrepancy
can be eliminated if we adopt a scenario where the observed GeV
$\gamma$--rays are produced by Comptonization of external photons (via
the ERC process), while the GeV SSC flux is well below the ERC emission, 
and thus ``hidden''.   In our 
analysis, we assumed that the the X--ray emission is produced via the SSC 
process, while the {\sl observed} GeV $\gamma$--rays are produced by 
the ERC process.  

In the context of this three-component scenario, we assume that 
the peak of the ``hidden'' SSC component can be still parameterized by 
$L_{SSC}$ and $\nu_{SSC}$.  We assume that $L_{SSC}$ and $\nu_{SSC}$ lie 
on or below the extrapolation of the {\sl ASCA} spectrum, but above 
the highest value of $\nu L(\nu)$
measured within the {\sl ASCA} band.  Since the spectra of QHBs generally have
$\alpha < 1$, the value of $\nu F (\nu)$ is the highest at the end of the
\asca bandpass (at 10 keV $\simeq$ 2$\times$10$^{18}$Hz) (see Fig.~3 in KTM98).
Further constraints arise from Eq. 2:  for an object with the observed 
$L_{sync}$ and $\nu{sync}$, a selection of $\delta$ sets a unique 
relationship between $L_{SSC}$ and $\nu_{SSC}$ such that 
$L_{SSC} = const \times (\nu_{SSC}/\delta)^2$.  
Since the VLBI data and $\gamma$--ray opacity arguments suggest that 
5 $< \delta <$ 20 for most blazars (e.g. Vermeulen \& Cohen 1994;  Dondi \&
Ghisellini 1995), we adopt the values of 5 and 20 as lower and upper limits
for $\delta$.  $L_{SSC}$ and $\nu_{SSC}$ are then constrained to be
inside the region defined from these conditions (for details, see KTM98).  
Once we obtain $L_{SSC}$, and $\nu_{SSC}$, we can calculate the strength 
of the magnetic field $B$ as follows:

\begin{equation}
B = 0.27 \left(\frac{R}{10^{-2}\mbox{pc}}\right)^{-1}
\left(\frac{\delta}{10}\right)^{-2} 
\sqrt{\left(\frac{L_{sync}}{10^{46}\mbox{erg/s}}\right)
\left(\frac{L_{sync}}{L_{SSC}}\right)}
\quad \mbox{(Gauss),} 
\end{equation}
and $\gamma_{peak}$ is then calculated using Eq. 1.

Using the size of the emitting region estimated from the observed time
variability via $R = c \Delta t \delta$ /  $(1+z)$, we infer $B$ to 
be 0.05 $\sim$ 1 Gauss (using $\delta$ = 10);  $B$ appears to be 
roughly similar for the two classes, but perhaps somewhat lower for HBLs 
than for QHBs.  
With these values of $B$, we estimate $\gamma_{peak}$ to be lower, 
10$^{3}$ -- $10^{4}$ for QHBs -- where the LE component peaks at lower 
frequencies ($\nu_{LE} \sim 10^{13}$ -- $10^{14}$ Hz) -- and higher, 
$\sim 10^{5}$ -- $10^{6}$ for HBLs, where $\nu_{LE}$ is at $10^{16}$ -- 
$10^{17}$ Hz.  This is illustrated in Fig.~3, which shows the correlation 
between $B$ and $\gamma_{el}$, where the HBLs preferentially occupy the 
right side of the diagram.  This picture is consistent with the fact that 
the TeV emission has been observed only from the HBLs which turned 
out to have higher $\gamma_{el}$.  This is because the energy conservation 
(or, equivalently, the Klein-Nishina limit) requires that Lorentz 
factors $\gamma$ of $\sim 10^6$ are the minimum necessary to produce TeV 
photons.  In the case of Mkn 421, the knowledge of $\gamma_{peak}$ allows 
us to calculate the energy of the ``seed'' low energy photons $h\nu_{seed}$ 
via the standard Compton formula 
$h\nu_{SSC} \simeq \gamma_{peak}^{2} \nu_{seed}$.  
For the TeV ($\sim 10^{25}$ Hz) radiation to be produced by Compton 
scattering, $h\nu_{seed}$ (as measured in the observer's frame;  we assumed 
$\delta = 10$) must be around 0.1 - 1 eV.  Interestingly, this means 
that the ``seed'' photons have {\sl lower} frequency than $\nu_{LE}$, 
where the energy density of the observed radiation spectrum peaks 
(around 0.1 -- 1 keV).  

As we argued above, for QHBs, where dense external radiation fields exist, 
the ERC emission most likely contributes more significantly than the SSC 
emission in the GeV $\gamma$--ray band (see, e. g., Sikora et al. 1997).  
This suggests that the difference of $\gamma_{peak}$ between QHBs and HBLs 
can well be due to  the larger total photon density in the jets of 
QHBs as compared with that of HBLs.  In any case, these simple estimates 
can be refined only by introducing a detailed model of the jet, 
including its dynamics as well as its surroundings.  

\begin{figure}[bt!]
\centerline{\psfig{file=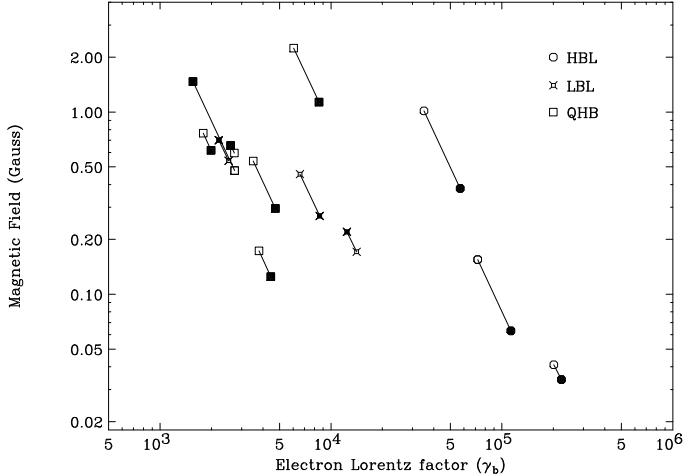,width=9cm,angle=90}}
\caption{Calculated magnetic field (B) versus Lorentz factor of
electrons radiating at the peak of the $\nu F\nu$ spectrum
($\gamma_{peak}$). In the calculation, we use the beaming factor
$\delta=10$. The size $R$ is estimated from the observed time
variability (open symbols; see KTM98). We also plot values calculated
with $R$ = 0.01 pc (filled symbols). }
\end{figure}

Recently, Ghisellini et al. (1998) 
performed detailed fitting based on the SSC and
ERC model to multi-frequency spectra and obtained the same conclusion
in that HBLs have higher $\gamma_{el}$ than QHBs.

\section{Multi-frequency Campaigns to Observe Mkn 421}

Among the GeV-emitting blazars, the BL Lac object Mkn 421 is unique as
the first -- and so far, the brightest -- member where the $\gamma$--ray
emission extends up to the TeV energies at a level allowing detailed
spectral and variability studies in the broadest range of wave bands.  
The simple continuum spectra of Mkn 421 from the radio to the UV and
X--ray bands obtained previously imply that the emission from the LE 
component is due to the distribution of charged particles radiating via the
synchrotron process (see, e.g., George, Warwick, \& Bromage 
1988).  At TeV energies, flares with variability time scale as short 
as 15 minutes have been observed (Gaidos et al. 1996). In order to 
avoid absorption of $\gamma$--rays due to pair production, this fast 
variability requires a beaming factor $\delta >$ 9.  

\subsection{1994 Observations}

Study of variability patterns and the spectral evolution across
different energy bands are the best (and perhaps the only) means to 
draw a general picture of non-thermal emission from blazars.  The 
first multi-frequency observation from radio to TeV $\gamma$--rays, 
conducted in 1994, revealed a TeV flare detected by the {\sl Whipple} 
Observatory (Kerrick et al. 1995), while the GeV $\gamma$--ray flux observed 
by {\sl EGRET} was nearly constant. The 24 hr {\sl ASCA}
observation, started one day after the onset of the TeV flare,
recorded a high level of 2 -- 10 keV X--ray flux peaking at 3.7
$\times$10$^{-10}$ ergs cm$^{-2}$ s$^{-1}$, a 10-fold increase over
the May 1993 value (Takahashi et al. 1996a). The flux level at
other wavelengths such as radio and UV was roughly the same as that
observed in the quiescent level (Macomb et al. 1995). The time
history of Mkn 421 obtained from {\sl ASCA} observation is shown in
Fig.~4, together with the change of the hardness ratio.  
The GeV flux measured by {\sl EGRET} (Macomb et al. 1995) 
and the flux above 500 GeV measured by {\sl Whipple} telescope
(Kerrick et al. 1995) are also shown in the figure.  The doubling time
scale is about a half day.

The 1994 {\sl ASCA} observation revealed that the X--ray flux 
variability in Mkn 421 is correlated with the spectral changes such that 
the spectrum is generally harder when the source is brighter.  In 
particular, the hard X--rays changed {\sl first}, followed by changes 
to the soft X--ray flux, implying a ``clockwise motion'' in the flux vs 
photon index plane (cf. Fig.~7).  Such behavior is typical at least 
for this particular BL Lac object and has been seen during earlier 
X--ray observations (cf. Tashiro 1992).  High sensitivity measurements by 
{\sl ASCA} allowed us for the first time to quantify the ``soft lag.''
We measure the lag in several energy bands as
compared to the 2 -- 7.5 keV band using the cross correlation function
of Edelson \& Krolik (1988).  The results are shown in
Fig.~5:  it appears that the hard X--ray variability leads that in
softer X--rays by about 1 hour.

With the knowledge of $\delta$ -- which we assume to be at least 5 -- 
the discovery of the soft X--ray lag allows us to calculate the
magnetic field from consideration of the synchrotron lifetime
(cooling) of the relativistic electrons.  The delay of the response of
the soft X--ray flux implies that this may be due to the electron
cooling effect, which is energy-dependent: the time $\tau_{sync}$ when an
electron loses a half of its energy would roughly be (in the
observer's frame) $5 \times 10^8$ $B^{-2}$ $\gamma_{el}^{-1}$
$\delta^{-1}$ s (cf. Rybicki \& Lightman 1979).  The peak observed
frequency of the synchrotron emission $\nu_s$ by an electron with
$\gamma_{el}$ is given as $\nu_s \simeq 1.2\times 10^{6}$ $B$
$\gamma_{el}^2$ $\delta$ Hz  for $z<<1$ (cf. Eq. 1).  If $E_{keV}$ is the
observed energy in keV, $\tau_{sync}$ is then expressed as $\tau_{sync}
\simeq 1.2 \times 10^{3}$ $B^{-3/2}$ $E_{keV}^{-1/2}$ $\delta^{-1/2}$ s.  
Since the lag is the difference of $\tau_{sync}$ at the different
X--ray energies, we obtain $\tau_{sync}$ $\sim$ 6000 s at 1 keV
(Fig.~5). If we take a beaming factor into account, $\tau_{sync}$ is
given as $2\times 10^{3} B^{-3/2} E_{keV}^{-1/2}\delta^{-1/2}$, and
$B$ becomes $\sim $ 0.2 Gauss for $\delta \sim 5$.  This, in turn,
yields $\gamma_{\rm el} \sim 5 \times 10^5 (E$/(1 keV)$^{1/2}$),
consistent with the value obtained by multi-frequency analysis
described above (also see Tavecchio, Maraschi, \& Ghisellini 1998)

Mastichiadis \& Kirk (1997) demonstrate that the flares of Mkn 421 are 
due to short-lived increase in the upper cutoff-energy of freshly
injected electrons, while keeping the electron energy distribution and
the magnetic field constant. Theoretical interpretation of the
clockwise motion has been discussed in Kirk et al. (1998) and Dermer (1998),
in terms of the balance between time scales of electron cooling and 
expansion of the emitting region.  

\begin{figure}[bt!]
\centerline{\psfig{figure=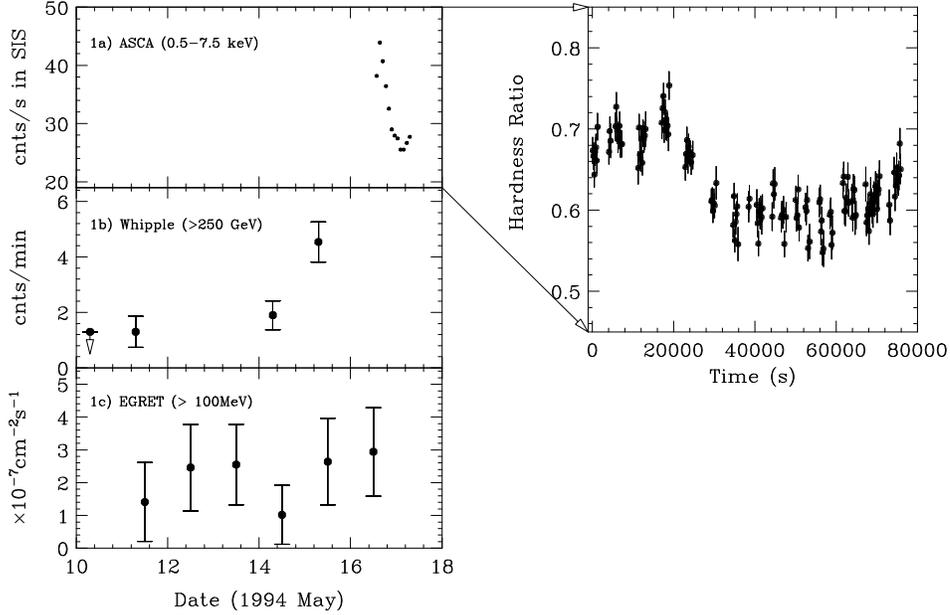,width=12.5cm,angle=90}}
\caption{Left panel: Time history of Mkn 421 emission obtained from
{\sl ASCA} (top), {\sl Whipple} Telescope (middle), and {\sl EGRET} 
(bottom) during the
May 1994 campaign. The {\sl ASCA} count rates are from the
summed SIS0 and SIS1 data. Right panel: hardness ratio (1.5-7.5 keV)/(0.5-1.5 keV).}  
\end{figure}

\begin{figure}[hbtp]
\centerline{\psfig{file=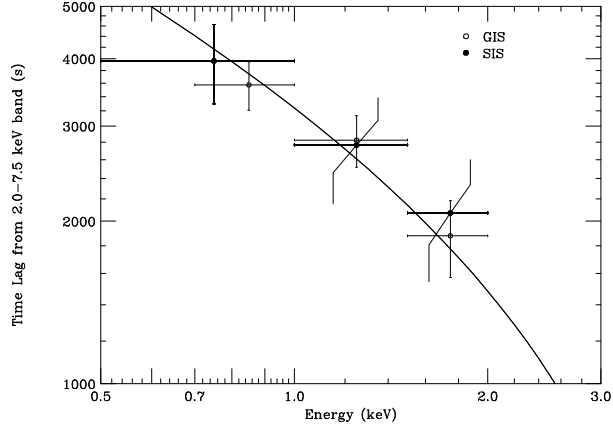,width=8cm}}
      \caption{
   Time lag of photons from Mkn 421 of various X--ray
        energies against the 2 - 7.5 keV band photons, calculated
        using the discrete correlation function (DCF) with data
        obtained in 1994 observation,
        The solid line corresponds to a
        fit with $\tau_{sync}(E) - \tau_{sync}(5{\rm keV}) =
        1.2 \times 10^{3}$ $B^{-3/2}$ $\delta^{-1/2}$
        $(E_{keV}^{-1/2} - (5)^{-1/2})$,
        which, for $\delta = 5$, yields $B \simeq 0.2$ Gauss, and
        $\tau_{sync}(1{\rm keV}) \simeq 6000$ sec (Takahashi et al. 1996a).  }
   \end{figure}

\subsection{1995 Observations}

In 1995, the X--ray observations started almost simultaneously with the
rise of the TeV flare (Buckley et al. 1996). In the campaign, short
observations of 8 -- 10 ks with \asca were spaced between 1 to 3 days
apart, such that it covered the TeV observations (Takahashi et al. 1996b).
Main aim was to monitor long term stability, rather than the short term
behavior we obtained in 1994.  The normalized light curve obtained from the
summed SIS0 and SIS1 data is shown in Fig.~6, together with TeV data
(Buckley et al. 1996), EUV data (Kartje et al. 1996) and optical data
(Wagner 1996). The variations shown on Fig.~6 are fairly similar in shape
in all four energy bands.  Although the peak-to peak amplitude is only 10
\% in the optical band, Mkn 421 shows variations which appear to be
correlated over an energy range of 13 decades of frequency, the widest
energy range over which such simultaneous changes have been observed so
far.  The X--ray emission does track (including the relative amplitude) the
general rise and decline seen in the TeV $\gamma$--rays.  It is clear,
however, that the variability of the X--ray flux has two components, one on
a $\sim$ one-day time scale as detected in the X--ray flare in 1994
(Takahashi et al. 1996a) and the other on a $\sim$ one-week time scale. The
best continuous coverage of the flare was obtained by \euve which resolved
the smooth rise and full of the flux, with variability of as much as a
factor of $\sim$1.5 over a span of $\sim$ 2 days (Kartje et al. 1997). 

\begin{figure}[bt!]
\centerline{\psfig{file=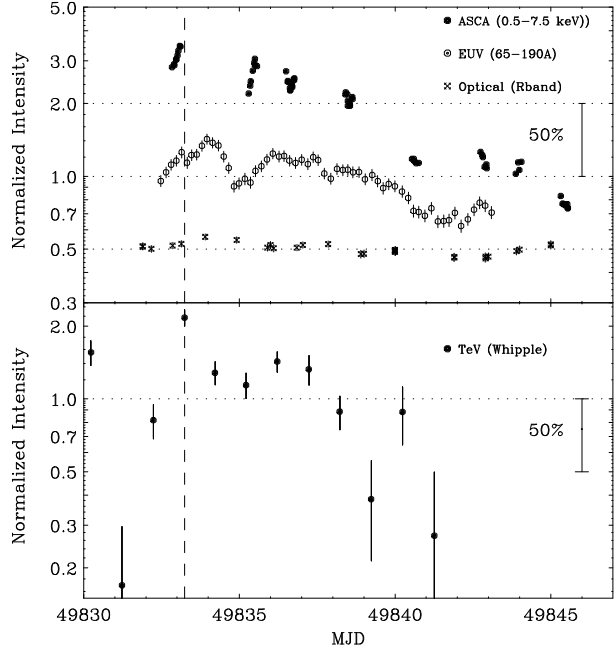,width=8cm,angle=90}}
\caption{Upper panel shows
Normalized X--ray (Takahashi et al. 1996b), extreme-UV (Kartje et al. 
1997), and optical (Wagner 1996) light curves for Mkn 421 taken
during the 1995 April-May multiwavelength campaign.  Lower panel shows
the normalized light curve of TeV $\gamma$--rays (Buckley et al. 1996).}
\end{figure}

The X--ray flux was at a similar level to that obtained in the
X--ray/TeV flare detected in 1994.  The X--ray flux
changed from 2.3 to 0.4 $\times 10^{-10}$ erg cm$^{-2}$ s$^{-1}$.  A
correlation between flux and the photon index as fitted to a simple
power law with free absorption is shown in Fig.~7 (which also shows the 
1994 data).  There is a general tendency, both in one-day
observation (1994) and two-weeks observation (1995), that the spectrum
becomes soft when the source is faint.  In both 1994 and 1995 observations, 
the X--ray spectra steepen towards higher energies and a single
power-law function and the absorbing column $N_{H}$ fixed at the
Galactic value does not fit any of the spectra well.  Instead, 
the peak frequency appears to shift from $\sim$ 1.4 keV to $\sim$ 
0.7 keV when the flux decreases (cf. Fig.~8).  
Again, there was a spectral variability within each
``sub-segment'', where the harder X--rays varied faster than the soft
X--rays. However, we could not obtain the time-lag between two energy
bands due to short observation time compared with the campaign in 1994.
Although the observations were not truly simultaneous, 
the amplitudes of the X--ray and TeV $\gamma$--ray flux variability were 
nearly the same during the decay portion of the observations.  

\begin{figure}[bt!]
\centerline{\psfig{figure=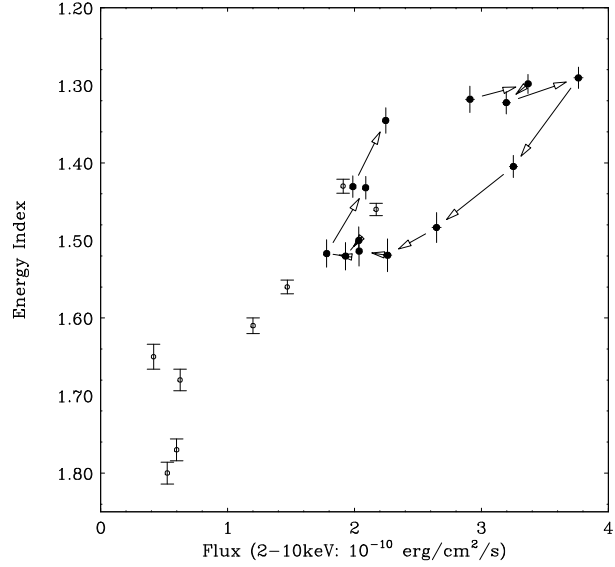,width=8cm,angle=90}}
\caption{ The evolution of the X--ray spectrum of Mkn 421 as a
function of the X--ray flux.  The model is a power law with free
absorption for data obtained in 1994 (filled circles) and 1995 (open
circles).}
\end{figure}

\begin{figure}[bt!]
\centerline{\psfig{figure=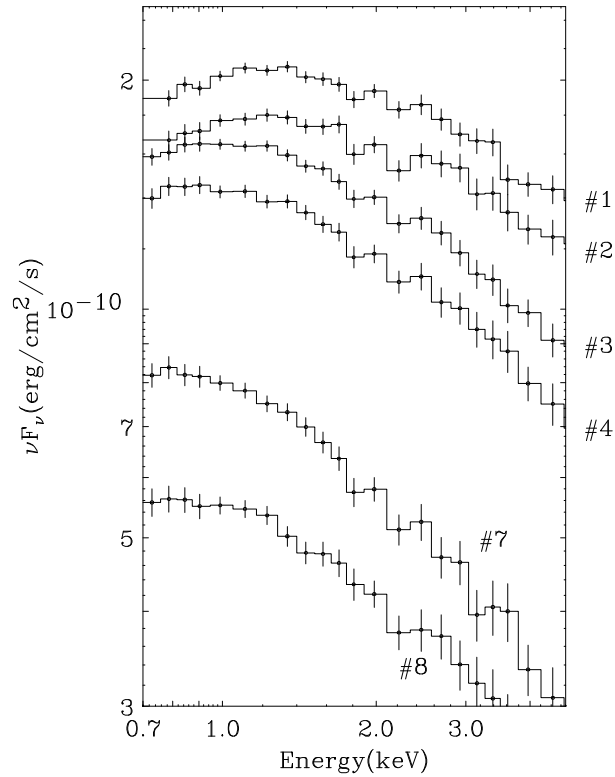,width=8cm,angle=90}}
\caption{The evolution of the X--ray spectrum in $\nu F \nu$ space for
data obtained in 1995. The shift of the peak frequency observed in the
declining phase during the 2 weeks campaign is clearly seen.}
\end{figure}

\subsection{1998 Campaign}

Judging by the results of previous campaigns which included TeV 
observations, correlations of inter-band variability of Mkn 421 
have proven to provide our best opportunity to understand the high energy 
emission from blazar jets.  In particular, the rapid intra-day 
variability seen in both X--ray and TeV energy bands gives us 
clues to study the physics of blazar jets.  However, the sparse sampling
of the previous campaigns has prevented us from obtaining definitive
conclusions. Inter-band correlations can only be confirmed
unambiguously if the observations are truly simultaneous and if they
extend for a period longer than several times the characteristic 
time scale.

With this in mind, we proposed an unprecedented seven-day continuous
observations with \asca (PI Takahashi), coordinated with EUVE (10 days
continuous; PI Takahashi), RXTE (low background orbits only; PI Madejski) and
SAX (two 60 ks observations, one is on April 21-22 and the other is
April 23-24; PI Chiappetti). At the same time, TeV detectors (CAT,
HEGRA, and Whipple), optical telescopes (coordinated by Mattox), and
22 GHz radio antennae (by Terasranta) attempted to observe the source
every night.  Here we present the preliminary results of the campaign;
the SAX results are presented elsewhere in these proceedings (cf. article by
Maraschi et al.).

\asca observation was performed during 1998 
April 23.97 -- 30.8 UT, yielding a net exposure of $\sim$ 280 ks.  The
time history of this observation obtained from the SIS detectors is
shown in Fig.~9b, together with that in the TeV (Fig.~9a), although the 
TeV observations are much sparser than those by \asca.  The
2 - 10 keV flux in the beginning of the observation was $1.2 \times
10^{-10}$ erg cm$^{-2}$ s$^{-1}$ and increased up to $5.0 \times 10^{-10}$ 
erg cm$^{-2}$ s$^{-1}$ at the maximum. More than 10 flares are clearly seen
superimposed on the general increasing trend. The doubling time scale 
of each flare is about 0.5 days. Continuous light curve
of 7 days implies that the source actually flares daily, and perhaps more 
often.  
Even though Mkn 421 is a bright X--ray emitter, this behavior is hard 
to study in any detail with the All Sky Monitor on board
RXTE:  our observation demonstrated that a large and sensitive
detector is indispensable for the study of variability in blazars.

The X--ray light curve contains multiple flares with time scale of
0.5 -- 1 day, superimposed on the general trend of gradually increasing flux.  
This is very similar to that seen in optical light curves of QHBs, 
where the high energy component peaks at GeV $\gamma$--ray energy
band (Wagner et al. 1996). This is consistent with the idea that the
largest amplitude of the rapid variability is caused by the highest 
energy end of the electron
distribution. The synchrotron emission due to these electrons
corresponds to eV range for QHBs and to keV range for HBLs such as Mkn 421.  
A good scenario would that either the ``blob'' of plasma
passed through the spatial region where shock are formed, or there is
some sort of standing shock in the jet, which is much larger than a
shock. Note that the particles can also escape upon crossing the source.  

\asca data clearly reveal spectral variability, seen as the time
variation of the hardness ratio in Fig.~9c. The 2 -- 7 keV energy index
ranges from 1.4 to 1.8. The comparison of the data from \asca, EUVE and 
RXTE indicates that the variability amplitudes in the LE (synchrotron) 
component are larger at higher photon energies. TeV light curve does show 
the same general trend, if the cross-calibration among the three participating 
TeV telescopes are accurate. Fig.~10 
shows the correlation between TeV flux by the Whipple telescope and
X--ray flux by \asca. It should be noted that the data were
obtained from truly simultaneous observations. In addition to the
correlation between 7-day \asca observation, one complete flare was
recorded by \sax and the Whipple telescope in the observation carried
out as a part of the 1998 campaign, but immediately preceding the \asca 
observation 
(Maraschi et al., these proceedings).  Detailed analysis of the 
evolution of the multi-frequency spectrum and shape of flares obtained 
from the campaign will be presented elsewhere (Takahashi et al. 1999, in 
preparation).

\begin{figure}[bt!]
\centerline{\psfig{figure=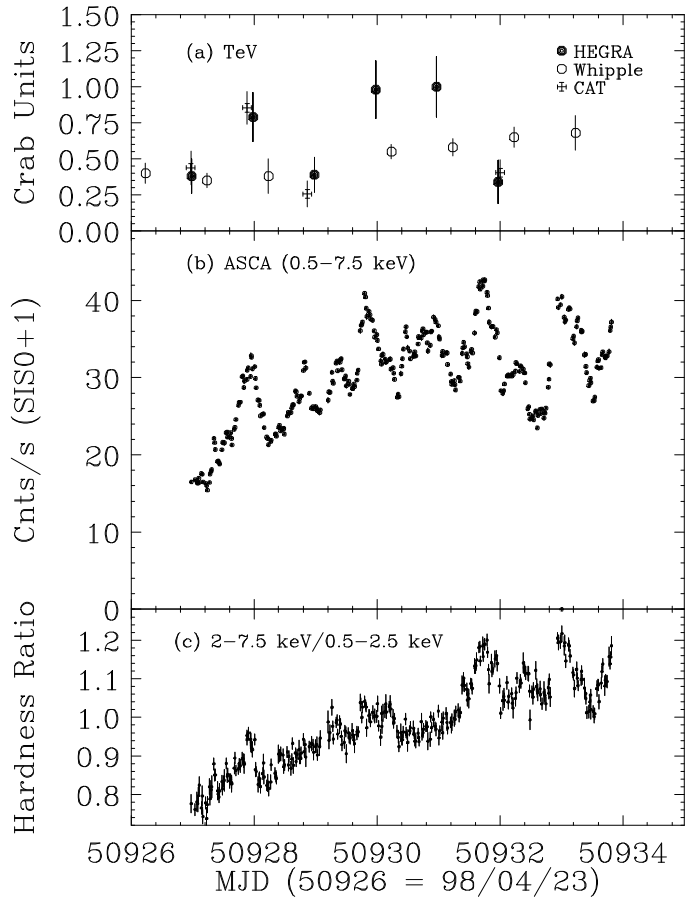,width=9cm,angle=90}}
\caption{1998 campaign for Mkn 421:  
(a) TeV light curves obtained by Whipple (Weekes et al. 1999, in 
preparation), HEGRA
(Aharonian et al. 1999, in preparation) and CAT (Degrange et al. 1999, in 
preparation) Cherenkov
telescopes.  (b) {\sl ASCA} light curve (summed SIS0 and
SIS1 data).  (c) Evolution of the hardness ratio measured 
in the \asca X--ray data (2.0 -- 7.5 keV)/(0.5 -- 2.0 keV).}
\end{figure}

\begin{figure}[bt!]
\centerline{\psfig{figure=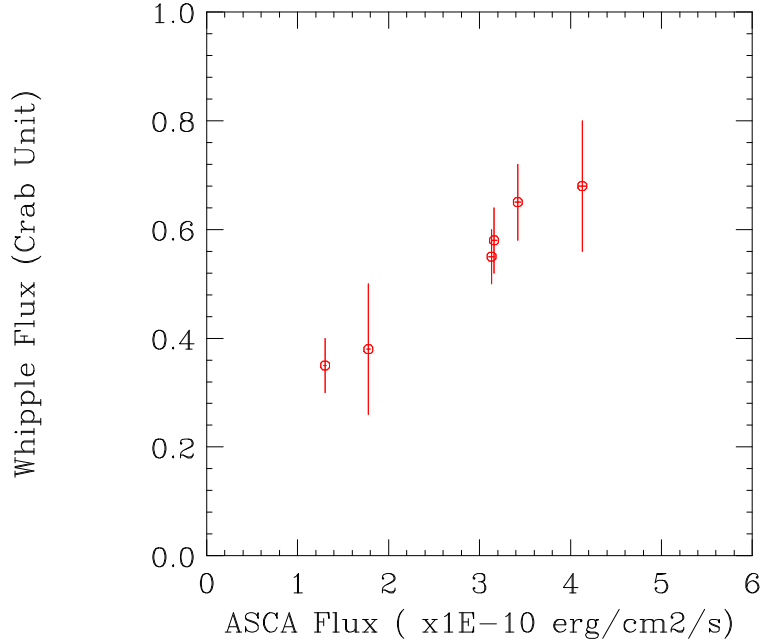,width=10cm,angle=90}}
\caption{Correlation between 2 -- 10 keV flux by \asca and the TeV 
$\gamma$--ray flux obtained by the Whipple telescope during 1998 
campaign. Observations are truly simultaneous.}
\end{figure}

\section{Summary and Conclusions}

The \asca data -- and, specifically, the spectral variability detected
with its sensitive instruments, in the context of the multi-frequency
analysis -- allow an investigation of the details of the emission in blazar 
jets, and in particular, 
an estimate of the physical parameters, such as $B$ and $\gamma_{el}$.
An application of the synchrotron self-Compton (SSC) model to the overall
spectral distribution and variability data for HBL-type blazars 
observed by \asca implies that the SSC model can explain all available data
quite well.  In particular, the values physical parameters of the radiating 
plasma inferred from the energetics requirement to produce the TeV 
$\gamma$--rays agree well with those inferred from the spectral variability 
observed in X--rays.  This model implies relativistic Doppler factors 
$\delta$ in the range of 5 - 20, consistent with the those derived from 
the VLBI data and from the limits inferred from $\gamma$--ray opacity 
to pair production, $\gamma\gamma\rightarrow e^{+}e^{-}$.  This is in 
contrast to QHBs, where the simple SSC models fail.  Instead, a 
three-component model, with the low energy peak due to synchrotron 
radiation, the X--ray emission produced by the SSC process, 
and MeV / GeV $\gamma$--ray emission due to Comptonization of external 
radiation photons, describes the observed data well.  

Previous multi-frequency campaigns of Mkn 421 with \asca and the
Whipple TeV telescope showed a very important connection between the
X--ray and the TeV energy bands. These observations are very
suggestive but not conclusive because of sparse sampling during flares
which occur on a time scale of $\sim 1/2 - 1$ day to one week. The great 
success of the big campaign conducted in 1998 is in providing valuable 
information about blazar jets via the measurement of spectral evolution 
of the emitted radiation, but, importantly, it also motivates the further
development of time dependent models for the structure of blazar jets, and 
acceleration of particles in these jets.  

\subsection*{Acknowledgements}

 The collaborators of the Mkn 421 campaign in 1998 include T. 
Takahashi, F. Aharonian, M. Catanese, L. Chiappetti, P. Coppi, B. 
Degrange, R. Edelson, H.Kubo, J. Kataoka, G. Madejski, F. Makino, H. 
Marshall, L. Maraschi, J. Mattox, E. Pian, F. Takahara, M. Tashiro, H. 
Terasranta, C. M. Urry, S. Wagner, and T. Weekes. We thank the CAT,
 HEGRA, and Whipple teams for providing with their data before
publication; and J. Kataoka and K. Yamaoka for the help of
analysis of $ASCA$ data.

\end{document}